\documentclass[aps,prl,twocolumn,superscriptaddress,nofootinbib,floatfix,
nobalancelastpage,nobibnotes]{revtex4}
\usepackage{epsfig}
\usepackage{graphicx}
\usepackage{caption}

\begin{document}
\title
{Fast flavor evolution in dense neutrino systems, as described in quantum field theory.}

\author{R. F. Sawyer} 

\affiliation{Department of Physics, University of California at
Santa Barbara, Santa Barbara, California 93106}

\begin{abstract}
 Investigations of dense neutrino cloud evolution through quantum kinetic equations led to the possibility of ``fast flavor" (FF) processes. It is shown here that the usual quantum kinetic equations, while signaling the instabilities that make 
some instances of FF possible, are being erroneously interpreted. Approaching the subject directly from the quantum field theory that defines the standard model shows the computational structures in most of the recent FF literature to be completely invalid. Our revisions also underlie what promise to be new early universe applications and a general result for relic supernova neutrinos that could be tested as a feature of the diffuse neutrino spectrum.
 \end{abstract}
\maketitle
\subsection{1. Introduction}
Fast neutrino-flavor evolution is defined by changes that take place on a timescale $T \sim[G_F n_\nu ]^{-1}$, where $G_F$ is the Fermi constant and
$n_\nu$ is the neutrino number density (units $\hbar=c=1$). It has become a major topic in discussions of neutrino clouds, and in turn of great interest in supernova calculations, and in early universe physics. It develops from an effective, nearly local coupling  among four neutrino fields induced by exchange of a virtual Z meson in the standard model. We omit all neutrino closed loop corrections (which produce factors $G_F$ with no accompanying $n_\nu$).

Two features of the neutrino-neutrino interaction in our systems that together make fast coherent behavior
 possible and calculable directly from very elementary quantum field theory (QFT) are:

1. That fast behavior (FF) leads to transformations over a distance of a fraction of a millimeter for $\nu$'s in an energy range of 1-20 MeV, when the number densities are in the range 1-1000 (MeV)$^3$.

2. That ``invariant T-matrix" amplitudes for $\nu+\nu \rightarrow \nu+\nu $,  $\bar\nu+\nu \rightarrow \bar\nu+\nu $, etc., in the absence of neutrino mass, are independent of the energies of the various neutrinos. 
The $\nu$'s have flavor indices as well, suppressed above.

In this environment, consider the period of a fast process: it is shorter than the $\nu$ scattering free-path in our application by a factor of 10$^{9}$ or more. When we concern ourselves with the nature of the process during that short time period, only a tiny fraction of the momenta of the individual neutrinos in the cloud can change at all. Momentum conservation is replaced by momentum preservation, a principle that must be held firmly in mind. We shall always embed our system in a periodic box big enough to follow our systems through a cycle of fast behavior. The initial angular distributions in the box 
are taken as absolutely constant in space.
We consider systems of neutrinos that begin in the pure flavor states in which they were produced by elementary processes.
Then the signals for the instability leading to FF change are the time-growing modes of an equation,
\begin{eqnarray}
i {\partial \over \partial t} \xi_j=\sum_k M_{j,k} \xi_k~, 
\label{general}
\end{eqnarray}
where $M_{j,k}$ is a non-Hermitian matrix that depends on the initial arrangement of flavors,
and the indices enumerate directions of flow. 

Those parts of the standard Sigl-Raffelt \cite{rs} equations (S-R) , or of related ``quantum kinetic equations" (QKE) that deal with the neutrino-neutrino interaction, cannot be an element in our conclusions, because at best they are applicable in systems that are dominated by neutrino scattering, whether from other $\nu$'s or other species. Here we are discussing another limiting case, where the sum in (\ref{general}) counts every one of the ($10^{22}$) neutrinos that are in the periodic box at the time of turn-on.  And we index each one by its  complete momentum, $\vec p_i$, not just its angle. The kernel of (\ref{general}) is of the order of the Fermi constant $G_F$. Taking the medium as translationally invariant over the short distances over which we follow the action, we do find fast transformations over an inverse time scale of $n_\nu$. These arise from (\ref{general}), but the path is to use QFT directly to obtain and understand these equations, which do not fit into the S-R. The steps to this conclusion involve a careful set of definitions for collective coordinates. Then we can not only address the absurdities of the QKE approach to this case but find some surprising results on final energy distributions.

The original S-R article does make the remark: ``We assume that the duration of one collision (the inverse of a typical energy transfer) is small relative to the inverse collision frequency.” But then the ``duration” is infinite for the processes that produce ``fast”, in which momenta are preserved. The literature that we are questioning ignores that fact.
 By contrast, we do not consider any collision whatever; we just transfer flavors between absolutely fixed momenta, as stated above, and also implicitly in (\ref{general}). 

There are also two reasons for invoking elementary quantum field theory:
 a) to obtain operator equations rather than equations for expectation values, enabling us to go one order of $\hbar$ higher than the S-R formalism.  This is needed in the consideration of seeding;
 b) to understand the role of an entirely new sector (discussed in section 3) coming from the ``u-channel" (in Mandelstam's old terminology \cite{mand}. In our momentum-preservation description it adds ``$\nu-\bar \nu$ mixing". This does not mean that we violate lepton number conservation, only that two single momentum rays can gradually exchange lepton number coherently.

One very simple example should suffice to show a drastic difference between the prevailing wisdom and our results. Take equal numbers of two flavors, in isotropic distributions, and suppose that the construction (\ref{general}) shows the growing instability. Then ask QKE practitioners what physics ensues. They answer ``none, because for each flavor (+1) there is an opposite (-1) in the initial state so that the total is zero already and nothing can happen."
One should just think about that hypothetical conclusion while putting it together with two features discussed above: a. The exact preservation of the neutrino momenta over our 1 mm./c instant of time; b. The fact that the invariant amplitudes are dependent only on flavor and angle, not on energies 
 The momentum vectors are trading flavors. But now we must address the likelihood that the energy spectra begin as strongly correlated to flavor, given the $\nu$ production processes in the supernova. 
When the momentum vectors assume new flavor values on a wholesale level, the previous flavor values necessarily get attached to new momenta which will be on-the-average higher for the lower set of energies, and on-the-average lower for the higher set. The reader can see the tendency toward energy equalization. Total macroscopic equalization will be discussed later. 

Use of the S-R equations did indeed provide the first evidence for the fast modes in well-described physical situations  \cite{fchir}-\cite{RFS4}, but most recent uses of them are incorrect to the core.  We cite here some examples, largely taken from \cite{richsen}, limiting to papers published in Phys. Rev., and Phys. Rev. Lett., mostly in the years 2021-2022 \cite{DS}-\cite{junk25}. There are many more in the pipeline. In any case, there exists also another sector of instabilities that is not addressed at all in present QKE's. It is discussed in section 3, and is probably destined to change all outcomes, whether or not one puts into effect the reforms called for in sec.2.  

 \subsection{2. Mechanics}
 To make our points we need to begin with a rather elaborate hierarchy of variables. 
 We consider $N_1$ individual $\nu$'s with momenta
 ${\bf p}_j$ where j runs through all 10$^{22}$ neutrinos in a ``beam" that is slightly diffuse in angle.
In the two-flavor ($\nu_e, \nu_x$) model underlying the simulations of the present paper we introduce
a Pauli matrix $\vec \sigma$-algebra (or, better, $\sigma_\pm, \sigma_3$), for this beam,
letting $a^j, b^j$ be the respective annihilators for $\nu_e$ and $\nu_x$ with momentum $\vec p_j$; 
 then defining  $\sigma_+^j =b_j^\dagger a_j $ for  $j=1$ to $N_1$,  and $\sigma_3^j =a_j^\dagger  a_j-  b_j^\dagger b_j$.

Intra-beam interactions being very strongly discouraged by the ubiquitous $(1-\cos\theta)$ factor,
we next discuss pairwise interactions between two beams where the second has the same number of $\nu$'s and uses 
$\vec \tau^j$, for for its description.

The standard model (Z exchange) neutrino-neutrino interaction local Hamiltonian, for our system, leaving out inconsequential terms with no flavor dependence is the basis for our work
We shall not here go into the complications ensuing in the three-flavor case with its eight operators replacing the three, except to say that our most important single result will be durable in this generalization.

Next we envision a set of different directions in space, with the ultimate aim to simulate the evolution of a 3-D 
ensemble of beams with some fairly small number of beams, say 60. We use a different Greek letter for each; in our work with 4 beams, we use 
$\vec \sigma,\vec \tau,\vec \xi,\vec \eta$.

The two-beam effective Hamiltonian written in terms of variables that are bilinears in $\nu$ creation and annihilation operators and that gives the S-R equations as its Heisenberg equations of motion is, 

 \begin{eqnarray}
&H_{\rm eff}={G_F\over {\rm Vol.} }\sum_{j,k}^{N_1}\Bigr [\sigma_+^j\tau_-^k +\sigma_-^j\tau_+^k 
+{1\over 2}\sigma_3^j \tau_3^k +
\bar \sigma_+^j \bar \tau_-^k +\bar\sigma^j_- \bar\tau^k_+
\nonumber\\
&+{1 \over 2} \bar \sigma^j_3 \bar\tau^k_3 
 -\bar \sigma^j_+\tau^k_- - \bar \sigma^j_-\tau^k_+ -\sigma_+^j\bar \tau_-^k -\sigma_-^j \bar \tau_+^k
 \nonumber\\
&-( \bar \sigma_3^j \tau_3^k -  \sigma_3^j \bar \tau_3^k )/2
\Bigr ] (1-\cos \theta_{\sigma, \tau})\, ,
\nonumber\\
\label{hammy}
\end{eqnarray}
with the obvious extensions to the cases in which we have more beams (additional greek letters and similar pairwise interactions involving all beams). The individual beams are to be of some small angular width, limited in principle by the precision sought in conjunction with
the angular factor $(1-\cos \theta_{\sigma,\tau})$. The number of beams in our list beginning with 
$\vec\sigma,\vec\tau.....$, as the first two, will need to be fairly large for good 3D simulations, as noted, but we can learn  much by studying one and two-dimensional simulations. Our computational methods come directly from using the Heisenberg equations for these variables, derived from their  commutators with $H_{\rm eff}$, and bearing in mind that the underlying momenta do not change in the course of a fast transition.

Next we deal with the intra-beam indices $j,k$ in the sum (\ref{hammy}), where these denote the individual $\nu$ rays that make up a beam, now in such a narrow cone of angles that that they do not generate a significant $(1-\cos\theta)$ factor. The number of rays in a beam is defined as $N_1$, and in the domains that we pursue could be $10^{22}$. We define collective $\sigma,\tau....$ operators, 
\begin{eqnarray}
\sum_j^{N_1} \vec \sigma^j \rightarrow \sqrt N_1 \vec \sigma~,~\sum_j^{N_1} \vec \tau^j \rightarrow \sqrt N_1 \vec \tau~.
\label{coll}
\end{eqnarray}
The redefined operators and their anti-particle mates for $ \bar \sigma, \bar \tau ...etc. $ obey the commutation rules of Pauli matrices. The factors of $\sqrt N_1$ that emerge 
combine with the ${\rm Vol.} ^{-1}$ factor in $H_{\rm eff}$ to replace the coupling constant in the Hamiltonian,
$G_F [{\rm Vol}]^{-1}$ by $G_F n$, where $n$ is the particle density in a single beam.
In the following, we choose time units such that this factor is unity.
\subsection{2.1 Two-beam example}
 We begin our serious dispute with the establishment by giving the complete details with respect to the two-beam instabilities.
 We take $\sigma$ as the right-moving amplitude and $\tau$ as the left moving amplitude.
 The Heisenberg equations of motion coming from commutators with (\ref{hammy}), now without the indices and with $G_F/{\rm Vol.}$ replaced by unity, are,
 \begin{eqnarray}
i( \partial/\partial t) \sigma_+=\sigma_3 \tau_+-\sigma_+\tau_3-\sigma_3\bar \tau_++\sigma_+ \bar \tau_3
 \nonumber\\
 i ( \partial/\partial t)  \tau_+=\tau_3 \sigma_+-\tau_+\sigma_3-\tau_3\bar \sigma_++\tau_+ \bar \sigma_3
  \nonumber\\
i( \partial/\partial t) \bar\sigma_+=\bar\sigma_3 \bar \tau_+-\bar\sigma_+\bar\tau_3-\bar\sigma_3 \tau_++\bar\sigma_+ \tau_3
 \nonumber\\
i ( \partial/\partial t) \bar \tau_+=\bar\tau_3 \bar\sigma_+-\bar\tau_+\bar \sigma_3-\bar \tau_3  \sigma_++\bar\tau_+  \sigma_3~.
\label{eom}
 \end{eqnarray}
 We consider initial states of pure flavor, ($\sigma_3,\tau_3,\bar \sigma_3,\bar\tau_3$),  each of which may be $\pm 1$, or zero if there is no corresponding beam at all.
We choose  ($1,0,0,-1)$ to obtain the model with initial configuration $\nu_e^R+\bar\nu_e^L$, chosen in refs. \cite{fchir}, and  \cite{IRT}, where at the half period time the state changes to  
 $\nu_x^L+\bar\nu_x^R$. Below we show how the instability enables this, giving 
 the complete transformation curve is of the form shown later in fig. 1 in sec. 4. Substituting into (\ref{eom}) linearizes the equations. We then find an eigenvalue of the matrix,\vspace{.1 in}

$\text{$m_1$}=\left(
\begin{array}{cccc}
 1 & 1 & 0 & -1 \\
 0 & -1 & 0 & 0 \\
 0 & 0 & -1 & 0 \\
 1 & 0 & -1 & 1 \\
\end{array}
\right);$
\hspace{1 cm}
\vspace{.1 in}

 with an imaginary part that signifies a growing mode
 $\exp[ t]$.  
 Why are we doing this when it is completely worked out in refs.
\cite{fchir}, \cite{IRT}? It is because we now want to ask all the authors that have bought into this work and its scores of sequels
a simple question: Suppose that we supply, in the initial state, counter beams in which the initial directions in the above set-up are swapped. 
An analytic approach  now requires that we look at 
8$\times$8 matrices rather than 4$\times$4's. They are of the block form,
\vspace{.2 in}

$\text{$m_2$}=\left(
\begin{array}{cccc}
 m_{1} & -m_1  \\
 -{~m_1}& m_1 \\
 \end{array}
\right);$
\hspace{.1 cm}
\vspace{.5 cm}
where the block $m_1$ is the 4$\times$4 matrix defined above.  Looking at the eigenvalues, it is trivial calculate the growth rate of the instability in this 8$\times$8 case. It is double that of the 4$\times$4 case.  

What is now happening in the medium, during our short period of time, is that in an ``equilibrium universe case" the flavor of each of our more than 10$^{22}$ rays in our tiny box is transforming itself through mixing in different flavor states, until at the half-period times the flavor values are the negatives of the beginning ones.

Now we look at the QKE approach to the same issues, beginning from eq. (1) in
\cite{junk25}
\begin{eqnarray}
i \partial_t \rho_{\bf n}(t)=[H,  \rho_{\bf n}]~,
\label{hoax}
\end{eqnarray}
where the descriptor for the evolution in a homogeneous medium is a c-number function $\rho_{\bf n}$ with $\bf{n}$ a direction of flow and $\rho$ (in the two flavor case) is a 2$\times$2 matrix in flavor space. Then in the two-beam case described above we would have one beam in the R direction and one in the L direction; and  
confirm from (\ref{hoax}) the results of ref.'s \cite{rs} and \cite{fchir} using the two clashing beams.

But then we added counter-beams in each direction and made them both equal combinations of the four flavor-lepton$^\#$ states (our above 8$\times$8 calculation).  We see that the QKE has no room for this, except to say
`` just add the flavors in all the right-moving beams, and separately in the left-moving beams, now giving $(0,0,0,0)$" in both cases. Nothing happens".

The first time we heard that we thought, ``let it pass, these guys don't care about all of those fast tradings of identity  
so long as the expectation of each flavor operator remains zero."  We did care much, as made clear
in papers directed at applications in cosmology \cite{RFS1} ,\cite{RFS2}. 
But the supernova people should care much as well, we now realize, because it is clear that fast trading of identities can equalize the \underline{energy} distributions of $\nu_e,\nu_x,\bar\nu_e, \bar\nu_x$, while leaving those flavor expectations fixed at zero, a topic to be discussed again in sec.2.3.

To reiterate with more specificity: there is a fatal flaw in the prevailing QKE approach to calculation.
When we said, ``take a initial flow with flavor parameters,$(\sigma_3, \tau_3,\bar \sigma_3,\bar \tau_3)$ given by $(1, 0,0, \pm 1)$
and then add a flow of equal intensity in the same directions with opposite parameters $(-1, 0,0, \mp 1)$",
we cannot assume that now those two flows are dormant because we obtained $(0, 0,0, 0)$. Indeed we showed
that, following all of the rules that produced the results of refs. \cite{fchir} and \cite{IRT} we obtained twice the original rate rather than the nullification expected in the QKE industry.
Of course, the ``nullification" could never have been the actual physical disappearance of the beams; it would have been with respect to their effects on the rest of the flow.

The caveat that summarizes this example is: When we begin with an assembly of pure flavor states we must start with separate beams for each flavor. 
\subsection{2.2 The same caveat, from a general argument.}
 A $\vec \sigma$ operator is built to be applied to composite states built of a very narrow-coned swarm of 
$N_1$ neutrino's of flavor A, say, as,
 \begin{eqnarray}
| \Psi_A\rangle= e^{i \sum_j \phi_j} \prod_{j= 1}^{N}| {\rm flav}_{j}^A \rangle \,,
\label{inter1}
\end{eqnarray}
where we explicitly put in the unknowable phase factor that comes with every $\nu$. 

 We expect this phase factor, which does not change during a period of coherent evolution, not to affect final results.   
 When we deal with the collectivization of the wave-functions within our very narrow cone in the way that matches
the corresponding collectivization, (\ref{hammy}), in $H_{\rm eff}$, we implement the step,
 
\begin{eqnarray}
 \prod_{j= 1}^{N}| {\rm flav_j^A}\rangle \rightarrow  |{\rm flav^A}\rangle~,
 \label{inter2}
 \end{eqnarray}
and the phase factor in (\ref{inter1}) then tags along mutiplying the whole expression.
Now suppose, e.g. that the states A and B are respective flavor choices for a right-moving $\nu_e$ beam where each $\nu$  carries $\sigma_3^j=1$, and a right-moving
$\nu_x$ with $\sigma_3^j=-1$, as in the ideal early-universe case,
 \begin{eqnarray}
| \Psi_B\rangle= e^{i \sum_j \phi_j'} \prod_{j= 1}^{N}| {\rm flav}_{j}^B \rangle \,.
\label{inter3}
\end{eqnarray}

Someone comes along and says ``but now we don't need so many beams, we can just add the two flavor variables together and use that as the initial flavor configuration for that part of the wave function."  giving,
\begin{eqnarray}
 | \Psi_A\rangle+| \Psi_B\rangle=e^{i \sum_j \phi_j} |{\rm flav^A} \rangle+e^{i \sum_j \phi'_j} |{\rm flav^B} \rangle
 \label{inter5} \,,
  \end{eqnarray}
  where we should have done,
  \begin{eqnarray}
 | \Psi_A\rangle| \Psi_B\rangle=e^{i \sum_j \phi_j} |{\rm flav^A} \rangle e^{i \sum_j \phi'_j} |{\rm flav^B} \rangle
 \label{inter6} \,.
  \end{eqnarray}
In view of its dependence on phase factors that are random, (\ref{inter5}) cannot play a role in a correct calculation. Our initial amplitude in the present problem must be in the form of a product over all 16 of the [flavor]$\times$[lepton number]  states, with no superpositions allowed \cite{RFS2}.

Our concern is with publications that are intended to make difficult systems like the supernova flow more computable by using some different basis than plane waves, for example angular moments. But it would appear that abandoning a basis of plane waves in favor of one of moments always means additive superpositions of plane wave states. In normal problems one can say ``If we use a complete set of angular functions that are best adapted to the geometry of the flow we can always get the plane-wave behavior back." In the present situation, though, we would have created something analogous to (\ref{inter5}) as an admissible state, and it would have the same fatal flaw as that state. 
Plane waves are the required basis for ``fast" $\nu$ work precisely because neutrinos move in exact straight lines during fast processes, no matter what the angular distribution might be.
\subsection{2.3 Energy spectrum equalizations} 
We consider possible equalization of energy spectra over all 6 participating flavor-lepton species. The argument is simple:
At that point at which we reach the half-way point in a complete oscillation, a single ray, $j$,
that began with a particular values $\Lambda^j=(\sigma^j_3,\tau^j_3,\bar\sigma^j_3,\bar\tau^j_3)$ has a totally different set of flavor values
 acquired by trades along the way, but exactly its original $|\vec p|$. It is just as true when $\sum_j \Lambda_j=0$. This leads to complete scrambling of energies, exactly because energies do not enter the dynamics, not because they do. 
 
 We stopped at the half cycle in formulating the above, for reasons we address more fully in sec. 4. In brief, our study of seeding, based on the next order terms in $\hbar$, raises serious questions about whether the fast process with periodic behavior at mean-field level actually extends beyond the first half cycle.
 \subsection{3. The u-channel, a fatal omission in the literature.}
This omission is ubiquitous, and not specific to the ``fast" sector. First: there are s,t,u channels in the basic Feynman graph that drives these effects (named after the $s,t,u$ variables of Mandelstam).
In ``s" a neutrino with four momentum $p_\mu^{(1)}$ and flavor $j$,``emits" a virtual Z, and becomes a neutrino with 
with four momentum $q_\mu^{(1)}$ and flavor $k$. The second incoming $\nu$ with four momentum $p_\mu^{(2)}$ and flavor $m$  ``absorbs" the virtual Z, and becomes a neutrino with four momentum $q_\mu^{(2)}$ and flavor $n$.
The s-channel flavor matrix is then $\delta_{j,k} \delta_{m,n} $ for the amplitude, where the indices take just two values $1,2$ for the flavors in the SU2 simplification.
Of course we get nothing interesting happening in the flavor space because of that  $\delta_{j,k} \delta_{m,n} $ dependence. 

To get the t-channel we exchange the flavors attached to the two four-vectors $q_\mu^{(1)}$ ,  $q_\mu^{(2)}$, making the change in the above: \newline $\delta_{j,k} \delta_{m,n}\rightarrow \delta_{j,n} \delta_{m,k}$. Now, imagining that we have backtracked a little to put in 
antiparticles correctly, and with a little work, we get exactly the Sigl-Raffelt equations from the last century \cite{rs} for the $\nu-\nu$ interaction.
But what about the ``u-channel"?  We get that from the s-channel amplitude for the case of incoming $\nu$ and $\bar \nu$  beams by exchanging the flavor and lepton number indices of $p_\mu^{(1)}$ and $q_\mu^{(2)}$, at the same time. Since that ``crosses" an incoming and outgoing Fermion line, lepton number mixing must be looked at carefully. But here is the fairly simple answer: The amplitude is a singlet, an invariant, in flavor SU2 (or SU3) space. And we are fairly sure people have noted that fact and said to themselves: ``so then nothing happens". 
But then one should realize that there is a fast u-channel $\nu-\bar\nu$ mixing interaction; though even to state the meaning
of what is happening one must remember that the ``beams" are all defined by a set of momenta that remain exactly the same during the beam interactions, according to the logic of the last section. It is the lepton numbers of the two beams that are traded.

This has been shown explicitly in excruciating detail in the ``supplementary material" to \cite{RFS3} for the case of two exactly parallel, $\cos \theta=1$ flows, one of $\nu$'s and one of $\bar\nu$'s. But it is easy to show generally that the complete angular dependence is  $(1+\cos \theta)$ for all u-channel effects, just as it is $(1-\cos \theta)$ for the t-channel effects.

We define operators $\vec \sigma$ for the transformation of a right-moving $\nu$ of any flavor into a right-moving $\bar \nu$ of the same flavor, where $\sigma_+=\bar a a^\dagger$, etc. Then we add an initial left moving $\bar \nu$ and follow the evolution.
This u-channel interaction is now by far the simplest of all ``fast" models, with the two beam case given by,
 \begin{eqnarray}
H_{\rm eff}^u=2 G_F n \Bigr[ \sigma_+\tau_ -+ \sigma_-\tau_+ \Big](1+\cos \theta)~,
\label{hambo}
\end{eqnarray}
the angle dependence now preferring beams in the same direction.
For the case of two beams in the same direction, taking time units such that $2 G_F n=1$, the eom are just,
\begin{eqnarray}
i \dot \sigma_+=\sigma_3 \tau_+~,
\nonumber\\
i \dot \tau_+=\tau_3 \sigma_+~,
\nonumber\\
i\dot \sigma_3=- \sigma_+\tau_- +\sigma_-\tau_+~.
\label{eom2}
\end{eqnarray}
A four-beam result is given in ref. \cite{RFS3} the two other beams $\xi$, $\zeta$, being at the common angle $\pi/2$.
Even leaving out these factors, e.g. at $\theta=\pi/2$, u-channel effects have important properties: 

1)The fast transitions 
by themselves will be four times as fast as those coming from  the more complex set of Sigl-Raffelt equations in the treatment of the t-channel effects.

 2)They work independently on the two (or three) ordinary flavors. 
 
 3) They establish that it is not true that fast transitions \underline{must} involve $\nu_e$ lepton number crossing (in the 2 flavor framework).

With the angular factors, the u-channel effects become much more the key process in the neutrino-sphere (or ``bulb") region of the SN, where the ordinary t-channel effects die off radically because of the narrowing of the angular distributions, but where the u-channel effects like nothing better. Furthermore the fact that $\bar \nu_e$ and $\nu_e $ spectral
\underline{shapes} get equalized (just the shapes-not the numbers) should be of great interest to designers of observations of the diffuse neutrino background, and could affect the design and priorities of experiments. 
However the process does require seeding.

\subsection{4. Seeding.} The ``u-channel" model of section 3, by virtue of its simplicity, provides an opportunity to explore ``seeding"  in a more definitive way than previously. More details are given in \cite{RFS3}, but we outline some essential points here. The u-channel instability of the last section that can drive neutrino-antlneutrino mixing is not activated in any variants of the neutrino mass-matrix and therefore must find initialization elsewhere if it is to prosper. 

In simple two and four-beam models  \cite{RFS3} we have shown how going beyond the mean-field approximation by one power of $\hbar$ provides an initiation mechanism. We define ``mean field approximation" (MFA) as simply that supplied by the $S-R$ equations, and where every quantity on the RHS is taken as an expectation in the medium.
 But our e.o.m as derived are completely operator equations; in these equations we choose to change the variables $\sigma, \tau$ of the last section to another set, 
 \begin{eqnarray}
&X=\sigma_+\bar\tau_- ~~, ~~Y=\sigma_- \sigma_+ + \bar\tau_+ \bar\tau_-   ~~,
\nonumber\\
&Z=(\sigma_3-\bar\tau_3)/2\,.
 \end{eqnarray}
We easily commute these products with $H^u_{\rm eff}$ of (\ref{hambo}) to get new e.o.m.'s
\begin{eqnarray}
 i \dot X=Z Y +Z^2/ N\,,~~{i \dot Y}=  2Z(X^\dagger- X)\,,
 \label{xtra}
\end{eqnarray}
\begin{eqnarray}
  {i \dot Z}=2(X-X^\dagger)\,.
  \label{eom3}
\end{eqnarray}

 The $Z^2$ term in (\ref {xtra}) comes from a second commutation to get operators into a correct order for
 expressing everything in terms of the new variables; implicitly it carries an additional power of $\hbar$ and provides the seed for something to happen. Of course now we wave the magic wand and declare that everything from now on will be taken as an expectation value. The above reasoning and outcome are closely related to those of an influential paper in condensed matter theory \cite{av}, but in very different language.

We need no explicit seeding to solve the new equations of development with pure lepton number initial values. Taking them as $\sigma_+=\tau_+=0$, $\sigma_3=-\tau_3=1$, leads to the results plotted as the solid curve for $\sigma_3$ in fig.1. We
refer to this as the ``quantum seeding case".
It is interesting that if instead we had used the lepton number changing seed $\sigma_1(t=0)=N_1^{-1/2}$ in the original mean-field equations we would have obtained an essentially perfect fit to this curve, as shown in fig. 1 (taken directly from \cite{RFS3}). However note that the fit is shown only out to the end of the second plateau.
 \begin{figure}[h] 
 \centering
\includegraphics[width=2.5 in]{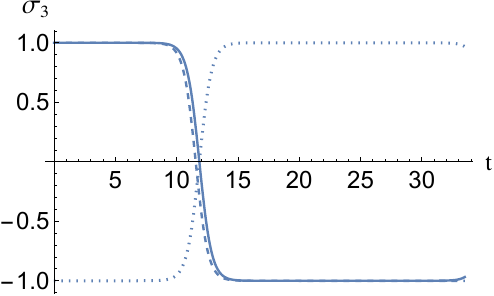}
 \caption{ \small } 
Dashed curve: Plot of the variable $\sigma_3=n_\nu-n_{\bar \nu}$ against time,  as determined by solution of the set (\ref{xtra})-(\ref{eom3}). The $n$'s are occupation averages in the beam that began as pure $\nu$. Dotted curve: the same for the beam that began as pure $\bar\nu$. Solid curve:
plot for the initially pure $\nu$ mode, but where we use the mean-field equations analogous to (\ref{eom2}), using a seeded
initial value as described in text .
 \label{fig.1}
\end{figure}
Here is the puzzle: the seeded mean-field case would go on periodically for a longer time, if allowed to, while the ``quantum seeded" result appears to develop a singularity that stops the computation exactly when we are expecting a second break that carries us back to the initial state. We believe that the probable answer is that the periodic behaviors are mean-field manifestations only, and that  FF events provide only a one-shot switch.  
 
After all, the reason that things in mean-field are periodic in the case of position independent flows has something to 
do with the fact that all of those classical gyroscopic pendulum solutions are producing specimens of Jacobian Elliptic Functions, already famous for their periodicities. 
But our form of seeding requires the next order in $\hbar$ beyond mean field and the JEF's are no longer the operative functions; nor can one recover that next order from one's semi-classical equations.  

In the supplementary material we show how additional evidence on this subject can come from simulations with small numbers (300+) of $\nu's$ and an unphysically large coupling. Our computations are marginal, but appear to support the above conclusion. 
We have been able to carry out the analogue of the argument based on  (\ref{xtra})-(\ref{eom3})  for a special example of the four beam case as well, but it is much more complex.
 In the supplemental material we add
 \cite{supa}, support from a different direction for this last conclusion. We also add material to describe some of the new early universe applications that are expected in our new framework, \cite{supb}, and speculate on the topic of how FF events might coexist with QKE solutions over much longer time periods \cite{supc}.
 
 We are indebted to Doug Singleton for an important discussion.
 
\section{Supplemental}
\subsection{a. Additional evidence against periodicity.}
As to the fate of our [ plateau$\rightarrow$sudden-change-of-sign $\rightarrow$ plateau scenario], in sec.4 we addressed the 
issues using our quantum-seeding formalism to conclude that coherence falls apart somewhere in that second plateau. And we concluded that the quantum flips are not part of a long term periodic oscillation.
We here offer more evidence from another direction. In 2003  Bell and Rawlinson, along with the present author,
 examined models as defined by (\ref{hambo}), but now truncated with a very small number, N, of $\nu$'s, and just solved the Schrodinger equation for the development of the system in time \cite{BRS}.  That is to say, in effect we turned up the coupling constant to an unphysical value so that the non-linear effects could be seen in small N systems. With our primitive laptops of the era, looking at exact solutions for particle numbers N in the 10's we found some evidence of fast mixing. Later Friedland and Lunardini pointed out that our interaction Hamiltonian lacked the essential $\sigma_3 \tau_3$ term, which, when included,
turned out to erase our result \cite{FL}.  But now, by total ironic coincidence, the exact model, with no such added term, is reborn in a different space, where the operators mix lepton with anti-lepton and leave flavors alone.
 
Therefore we now look again at the $N_1$ dependence of a $\sigma_3$ plot beginning at unity, where $\sigma_3$ now measures (leptons-antileptons). For $N\sim10$'s, we saw evidence for ``fast" ; now in the early $N\sim100$'s we see the initial plateau forming, along with a later fairly fast downturn; at 300 things sharpen and we see flattening as we approach what we anticipate is another plateau, this time near $\sigma_3=-1$. It's just the half-cycle. We do not believe that it is tending towards a revival into the full cycle. Thus the study of seeding in the main text, and independently looking at analogues of old work with Bell and Rawlinson, extended to larger N, and hinted at by Roggero \cite{rog} tell you that we get only to the first half-period, after which the system dissolves back into something very like its everything-flipped state, \underline{except} for the under-the-table redistribution of energies!

\subsection{b. The nature of the neutrino content of the early universe in the temperature range, $1<T<10 $ Mev} 
In this system, do we have flavor near-equilibrium that changes only slowly, since the thermal electrons and positrons barely have begun getting rid of themselves in favor of photons?  
Or is there rapid flavor mixing that can lead to important physical consequence?
 Our answer, as given in \cite{RFS1}, is the latter.  
 For simplicity we argue this while putting aside for the moment the lepton-number oscillations described in section 3, above.  Just to summarize the basics we start in the most elementary configuration as an initial state: two beams clashing head-on, each an incoherent equal mixture of  $\nu_e,\bar\nu_e,\nu_x, \bar\nu_x$ states. To follow the flavor, we define the bilinear operators $\sigma_+$, $\bar \sigma_+$ for the up-moving stream, and 
 $\tau_+$, $\bar \tau_+$ for the down-moving stream, where these are the operators that change flavor from -1 to 1. The operators that measure flavor are $\sigma_3$, $\bar \sigma_3$ for the up-moving stream, and $\tau_3$, $\bar \tau_3$ for the down-moving. The Hermitian conjugates, $\sigma_-$ etc. follow along in a totally dependent way, and do not require their own equations. 
 
Now suppose that after deriving the 8 non-linear equations from commutators with the effective Hamiltonian obtained by a heavy Z exchange, we take initial values of zero for the flavor measuring operators, 
$(\sigma_3, \bar \sigma_3, \tau_3, \bar \tau_3)$
 based on the perceived cancellation between the $\nu_e$ and $\nu_x$ parts.  The short term development of the system as given by the four equations for $(\sigma_+, \bar \sigma_+ , \tau_+, \bar \tau_+)$ is trivially that nothing happens to order t.  

But this is not the end of the story.
Taking $\sigma_3=0$ because we have equal initial numbers of up-moving $\nu_e$ and $\nu_x$ is making exactly the mistake denounced in sections \#1 and \#2, namely, adding initial flavors within a beam. We repeat our litany about momentum-preservation as opposed to ``quantum kinetic equation" yet again for this specific question. In the QKE conception a beam begins in a state that is a narrow angular bundle of $10^{22}$ $\nu$'s, but it can't do the book-keeping to know who started in one or another flavor state;  in our approach we simply accept the necessity of having  separate beams for each flavor value in the S-R equations.

At a minimum, then, we need two separate right-going beams for our two initial flavors; and separately the same for the left going beams and all the corresponding anti-particle beams. We now have four times as many equations as before, making 32 in all. Again only one half of these enter into the linearized instability matrix. The complete 16 dimensional response matrix analogous to is available to interested parties at the authors e-mail address. The reference for the underlying equations themselves is \cite{RFS1} . 

The instability provides exactly our answer to the rhetorical question asked and answered at the beginning of this section. If we look at the early universe at the level of the single neutrino, as defined by its
momentum state, over a time period much, much shorter than the scattering time, then we see a turmoil of 100\% flavor oscillations ensuing when it begins in a pure flavor state. There are cosmological consequences of this; first in the fact it can provide a free source of energy in the KeV region that allows the simplest Dodelson-Widrow models \cite{DW} with the ability to use bilinear $\nu, \nu_{\rm sterile}$ couplings that are many orders of magnitude less than previously thought; second, in what could be a substantial reworking of the calculations, e.g. by Dolgov et al., in the region in which the positrons disappear
\cite{dol}, and one struggles to find the final photon and neutrino distributions. 

\subsection{c. Combining  FF processes and ``quantum kinetic equations". }
Nowhere in the above have I stated or implied that QKE's are unimportant in supernova calculations, only that they are the wrong container for FF effects that take place in a region that is of order $(nG_F)^{-1}$ in size. 

How can we find a consistent way of using each in its domain of validity? Some implications of our seeding discussion in the previous  section may be useful here. A single FF cell is very localized in space and limited in time duration. In the FF equations the space derivatives do not enter at all. Nor do neutrino energies.
In each of our countless little boxes, everyone in the domain interacts with everyone else, and kinetic energies are heedlessly traded among different flavors and different directions.  

This should lead to nearly exact equalization of energy spectra across the full six species (flavors $\times$ lepton-ness). As stressed earlier, this is not because energies enter the governing equations for fast phenomena; it is because they do not. All of the explicit energy trading is out of sight of the FF equations, since the energies do not even appear in $H_{\rm eff}$.

Influential precursers to the explosion of work on fast processes are found in the big literature in the ``pendulum in flavor space" ,  which began as a fascinating exercise in which, e.g., segments of the $\nu_e$ and $\nu_\mu$ energy spectra could be swapped in ways that were surprisingly fast and that depended strongly on the $\nu$ mass matrix  \cite{DFCQ} ,\cite{HRSW}. That is to say: in our present terms it relied on a particular way of seeding, one that appeared inevitable for that era. 

But our results argue strongly against that swap, and in favor of rapid spectral equalization.
Furthermore, by now, as noted in sec. 5, it has been shown in two ways how, in the absence of neutrino mass, effective seeds will be provided in the simplest models; both are higher order in $\hbar$ however.  
Indeed, because the mainstream vein of literature is all rooted in semi-classical approximations from the outset, it cannot supply the order $\hbar$ corrections for seeding (beyond, or instead of the order $\hbar$ neutrino mass matrix),

A recent ``pendulum" paper \cite{junk25} is worthy of note in other respects.  By considering only the 
integral of the amplitude over all energies, so that the dependence is completely on angles, and then assuming azimuthal symmetry, so that the distribution is over polar angle only, it is hoped that the equations may now be simple enough for  answers on criteria for instability. Solutions are shown that are based on particular assumptions for the mass matrix. But we see that the azimuthal superposition at the foundation of this work
violates the prohibitions established in sec.1 and sec.2.
And the model itself is intrinsically unable to track the energy spectra. Even if it could be rescued in multi-beam form, it could not confirm or deny our most important prediction; namely the equalization of these spectra.

Of course, when on their trip outwards in the supernova, $\nu$'s enter a domain in which FF fades and scattering takes over, this by itself will reintroduce  some energy spectrum differences.  But they are far from the ones that would have emerged from the inner core were it not for the energy trades in the FF events.
It appears therefore that the spectral homogeneity of all six neutrino species can be checked in measurements of the diffuse neutrino spectrum that are said to be in the works.  
 
In our results the flavor turnover plot has a universal shape, for large values of N, and with no appreciable seeding from neutrino masses. 
 First there is a plateau of time extent proportional to $[\log N ]G_F n$ at an initial value that has been scaled to unity, then a sudden drop to very nearly $(-1)$, almost exactly reversing the initial flavor value, then another plateau--until? Now the interesting part surfaces: in the pendulum calculations, and in any seeded mean-field calculations, it is periodic. But when we replace ordinary seeding by a quantum-correction-seeding it appears that it is not able to continue on its periodic course. 
  
 Therefore it seems possible that the early universe neutrino content remains almost the thermal equilibrium mix, but with this enormous rate of little localized blips that can do some physics on the side, as suggested in refs. \cite{RFS1} and \cite{RFS2}. In our immediate pre-electron-annihilation era we estimate that $\log N$ is about 40, meaning that there should be big flavor mixing (during the blip) about 2\% of the time.


\subsection{Bibliography}
\begin{thebibliography}{99}
 \bibitem{rs} G. Sigl and G. Raffelt, Nucl. Phys. {\bf B406}, 423 (1993)
 \bibitem{mand} S. Mandelstam, Phys. Rev. {\bf 112}, 1344 (1958); Phys. Rev. {\bf 115}, 1741 (1959)
\bibitem{fchir} S. Chakraborty, R. S. Hansen, I. Izaguirre, G. Raffelt, Nucl. Phys.{\bf B908}, 366 (2016), arXiv:1602.00698
\bibitem{IRT}I. Izaguirre, G. Raffelt, I Tamborra, Phys. Rev. Lett. {\bf 118}, 021101 (2017), arXiv:1610.01612
 \bibitem{RFS4} R. F. Sawyer, Phys. Rev. Lett. {\bf 116}, 081101 (2016), arXiv:1509.03323  
\bibitem{richsen} S. Richers and M. Sen, arXiv:2207.03561
\bibitem{f1}A. Dighe and M. Sen, Phys. Rev. {\bf D97}, 043011 (2018), arXiv:1709.06858
\bibitem{f2}B. Dasgupta and M. Sen, Phys. Rev. {\bf D97}, 023017 (2018),  arXiv:1709.08671
\bibitem{f3}S. Abbar and H. Duan, Phys. Rev. {\bf D98}, 043014 (2018),  arXiv:1712.07013
\bibitem{f6} C. Yi, L. Ma, J. D. Martin, and H. Duan, Phys. Rev. {\bf D99}, 063005 (2019), arXiv: 1809.09137
\bibitem{f9} R. Glas , H.-T. Janka , F. Capozzi , Manibrata Sen, B. Dasgupta , A. Mirizzi, G. Sigl,
Phys. Rev. {\bf D101}, 063001 (2020), arXiv:1912.00274

\bibitem{junk1} L. Johns and H. Nagakura, Phys. Rev. {\bf D103}, 123012 (2021) 
\bibitem{junk2} Sherwood Richers, Don E. Willcox, Nicole M. Ford, and Andrew Myers,
Phys. Rev. {\bf D103}, 083013 (2021) 
 \bibitem{bd2 }S. Bhattacharyya and B. Dasgupta,   Phys. Rev. {\bf D102} , 063018 (2020) 
\bibitem{junk3}Sajad Abbar, Francesco Capozzi, Robert Glas, H.-Thomas Janka, and Irene Tamborra,
Phys. Rev. {\bf D103}, 063033 (2021)
\bibitem{junk4}Francesco Capozzi, Sajad Abbar, Robert Bollig, and H.-Thomas Janka
Phys. Rev. {\bf D103}, 063013 (2021) 
\bibitem{junk5}Joshua D. Martin, J. Carlson, Vincenzo Cirigliano, and Huaiyu Duan
Phys. Rev. {\bf D103},  063001 (2021)
\bibitem{junk6}Soumya Bhattacharyya and Basudeb Dasgupta
Phys. Rev. Lett. 126, 061302 (2021)
\bibitem{junk7}Lucas Johns, Hiroki Nagakura, George M. Fuller, and Adam Burrows,
Phys. Rev. {\bf D102}, 103017 (2020)
\bibitem{junk8}Manu George, Meng-Ru Wu, Irene Tamborra, R. Ardevol-Pulpillo, and H.-T Janka, Phys. Rev. {\bf D102}, 103015 (2020)
\bibitem{junk9}Robert Glas, H.-Thomas Janka, Francesco Capozzi, Manibrata Sen, Basudeb Dasgupta, Alessandro Mirizzi, and Günter Sigl,Phys. Rev. {\bf D101}, 063001 (2020)
\bibitem{junk11}S. Abbar, H. Duan, K. Sumiyoshi, T. Takiwaki, and M. C. Volpe, Phys. Rev. {\bf D101}, 043016 (2020) 
 \bibitem{junk12} S. Abbar, F. Capozzi, R. Glas, Phys. Rev. {\bf D103},063033(2021) 
  \bibitem{junk13} Bhattacharyya, Soumya; Dasgupta, Basudeb , Phys. Rev. D106,103039 (2022)
 \bibitem{junk15}F. Capozzi, M. Chakraborty, S. Chakraborty, and M. Sen, Phys. Rev. Lett.125,251801 (2020)
 \bibitem{junk16} F. Capozzi, M. Chakraborty, S. Chakraborty, Phys. Rev. {\bf D106}, 083011 (2022)
\bibitem{junk17}F. Capozzi,  S. Abbar, R. Bollig, Phys. Rev. {\bf D103},3013 (2021)
\bibitem{junk18}S. Abbar, H. Duan, K.  Sumiyoshi et al.,Phys. Rev. {\bf D104}, 043016 (2021)
\bibitem{junk19}H. Duan, J.D. Martin, S. Omanakuttan,
Phys. Rev. {\bf D101}, 123026 (2020)

\bibitem{junk20}M. George,M.-R. Wu, I. Tamborra , R. Ardevol-Pulpillo, and H.-T. Janka
Phys. Rev. {\bf D102},  103015 (2020)

\bibitem{junk21}R. Hansen, S. L. Rasmus, S. Shalgar,I. Tamborra,
Phys. Rev {\bf D105} , 123003 (2022)

\bibitem{junk22} O. Just,  S. Abbar,  M.-R. Wu, I. Tamborra, H.-T. Janka, and F. Capozzi,
Phys. Rev. {\bf D105},    083024 (2022)

\bibitem{junk23}J. Martin,J. Carlson,V. Cirigliano, and H. Duan,
Phys. Rev. {\bf D103}, 063001 (2021)

\bibitem{junk24}H. Nagakura, A. Burrows, L. Johns, and G. M. Fuller,
Phys Rev. {\bf D104}, 083025 (2021)

\bibitem{junk26}S. Richers, D. E. Willcox, N. M. Ford, 
Phys. Rev. {\bf D104}, 103023 (2021)
\bibitem{junk27}S. Richers, H.  Duan,  Meng-Ru Wu, S. Bhattacharyya, M. Zaizen, M. George, Chun-Yu Lin, and Zewei Xiong
 Phys. Rev. {\bf D106}, 043011 (2022)

\bibitem{junk28} S. Shalgar, I. Tamborra,
Phys Rev. {\bf D105},  043018 (2022)
\bibitem{junk29}M. Zaizen, T. Morinaga, 
 Phys. Rev. {\bf D104}, 083035  ( 2021)
 \bibitem{junk25}I. Padilla-Gay, I Tamborra , G. Raffelt, 
Phys. Rev. Lett., 128, 121102 (2022)
 \bibitem{RFS1} R. F. Sawyer, arXiv: 2104.02771
\bibitem{RFS2} R. F. Sawyer, arXiv: 2111.07204
\bibitem{RFS3} R. F. Sawyer, arXiv: 2206.09290
\bibitem{av} A. Vardi, J. R. Anglin, Phys. Rev. Lett. {\bf 86}, 568 (2001), arXiv:physics/0007054
\bibitem{BRS} Nicole F. Bell, Andrew A. Rawlinson, R. F. Sawyer, Phys.Lett.{\bf B573}, 86  (2003), arXiv:hep-ph/0304082 
 \bibitem{FL}Alexander Friedland , Cecilia Lunardini, Phys.Rev. {\bf D68},  013007 (2003), arXiv:hep-ph/0304055
\bibitem{rog}Alessandro Roggero, Phys. Rev. {\bf D104} (2012), arXiv2103,11497


\bibitem{supa} See Supplemental Material at [ ]   for the alternative work on quantum seeds.
     \bibitem{DW} S.Dodelson and L. M. Widrow, Phys. Rev. Lett. {\bf 72} 17, (1994), arXiv:hep-ph/9303287
\bibitem{dol}A.D. Dolgov, S.H. Hansen, D.V. Semikoz, Nucl.Phys. {\bf B503} ,426 (1997)
arXiv:hep-ph/970.3315
  \bibitem{supb} See Supplemental Material at [ ]   for the possible applications to the early universe.
\bibitem{DFCQ}H. Duan, G. M. Fuller, J. Carlson, and Y.-Z. Qian,  Phys. Rev. D 74, 105014 (2006).
\bibitem{HRSW} Hannestad, G. G. Raffelt, G. Sigl, and Y. Y. Y. Wong, Phys. Rev. D 74, 105010 (2006); Phys. Rev. D 76, 029901 (2007).
\bibitem{supc}See Supplemental Material at [ ]  for speculation on coexistence with QKE's.
\end{thebibliography}
\end{document}